\documentclass[12pt,preprint2]{aastex6}

\newcommand{\Ha}{H$\alpha$}

\newcommand{\Htwo}{H\textsubscript{2}}

\newcommand{\kms}{km~s{$^{-1}$}}

\newcommand{\cmq}{cm$^{-3}$}

{\Large {\Large }}

\newcommand{\oiii}{[\ion{O}{3}]}

\newcommand{\nii}{[\ion{N}{2}]}

\newcommand{\Vd}{V$\rm_{3-D}$}

\newcommand{\Vlowoiii}{V$\rm _{low,[O III]}$}

\newcommand{\Vnii}{V$\rm_{[N~II]}$}
\newcommand{\Voiii}{V$\rm_{[O~III]}$}

\newcommand{\Vr}{V$\rm_{r}$}
\newcommand{\Vrn}{V$\rm_{r,[N~II]}$}

\newcommand{\Vt}{V$\rm_{tan}$}
\newcommand{\Vtnii}{V$\rm_{tan,[N~II]}$}
\newcommand{\Vtoiii}{V$\rm_{tan,[O~III]}$}
\newcommand{\PAnii}{PA$\rm_{[N~II]}$}
\newcommand{\PAoiii}{PA$\rm_{[O~III]}$}

\newcommand{\mol}{N$\rm _{2}$H$\rm ^{+}$}

\newcommand{\crossing}{Crossing}
\newcommand{\cloud}{Orion-S Cloud}
\newcommand{\nil}{NIL}
\newcommand{\core}{Core}
\newcommand{\Vlongoiii}{V$\rm _{long,[O~III]}$}
\newcommand{\Vshortoiii}{V$\rm _{short,[O~III]}$}
\newcommand{\Slongoiii}{S$\rm _{long,[O~III]}$}
\newcommand{\Sshortoiii}{S$\rm _{short,[O~III]}$}
\newcommand{\ledge}{Ledge}
\newcommand{\extledge}{Extended-Ledge}
\newcommand{\midGrp}{Middle-Group}
\newcommand{\CrGrp}{Crossing-Group}
\newcommand{\east}{East}
\newcommand{\west}{West}
\newcommand{\coup}{COUP~632}

\newcommand{\mm}{mm~9}
\newcommand{\drs}{DRS 1186}

\begin{document}

\title{Deciphering the 3-D Orion Nebula-IV: The HH~269 flow emerges from the Orion-S Embedded Molecular Cloud}

\author{C. R. O'Dell\affil{1}}
\affil{Department of Physics and Astronomy, Vanderbilt University, Nashville, TN 37235-1807}

\author{N. P. Abel\affil{2}}
\affil{MCGP Department, University of Cincinnati, Clermont College, Batavia, OH, 45103}

\and

\author{G. J. Ferland\affil{3}}
\affil{Department of Physics and Astronomy, University of Kentucky, Lexington, KY 40506}

\begin{abstract}
We have extended the membership and determined the 3-D structure of 
the large (0.19 pc) HH~269 sequence of shocks in the Orion Nebula. 
All of the components lie along a track that is highly tilted to the plane-of-the-sky and emerge from within the Orion-S  embedded molecular cloud. 
Their source is probably either the highly obscured mm 9 source associated with a high N2H+ density core (more likely) or the more distant star COUP 632 (less likely). The former must be located in the Photon Dominated Region (PDR) underlying the ionized surface of the Orion South Cloud, while the latter would be embedded within the cloud.
 The flows seem to be episodic, with intervals of 1900 to 2600 years or 700 to 2600 years if COUP 632 is the source. 
\end{abstract}

 \section{Introduction}
 \label{sec:Intro}
 This paper is the fourth in a series reporting on properties of features within the Orion Nebula. PaperI \citep{ode20a} dealt with large-scale features, especially the foreground layer of ionized material, the placement of the embedded Orion-S Molecular Cloud (henceforth the \cloud), and the more distant overlying ionized layer designated as the nearer ionized layer (the \nil ).
 Paper II \citep{ode20b} identified a new major feature of the nebula lying ESE--WNW across nebula and with the \cloud\ on its northern boundary. These first two papers used spectra averaged over blocks of 10\arcsec $\times$10\arcsec\ (velocity resolution of 10 \kms).
 Paper III \citep{ode20c} used several full spatial resolution (about 2\arcsec) sequences of spectra, denoted therein as Profiles, to determine the properties of the brightest part of the Orion Nebula that occurs at the NE boundary of the \cloud . In that study a central 30\arcsec\ diameter region designated  as the \crossing\ was spectroscopically very complex and informed the spatial structure of the photoionized layer of the \cloud\ that faces the observer.  These three papers give summaries of our preceding knowledge of the structure of the Huygens Region (the bright central portion of the Orion Nebula) and the progressive improvement of this knowledge.

From a star formation point of view, the arguably most interesting region identified in Papers I--III is the \cloud . These earlier papers established that this feature is a host for young stars, many of which have collimated outflows. The study of these outflows requires using images and spectra in a different manner.
 The most important procedural differences are that in the present study we draw on time-lapse images made with the Hubble Space Telescope to identify high velocity flows associated with HH~269 and that we employ high spatial resolution radial velocities of these features to determine their 3-D locations and motions. This then allows us to precisely constrain the location of the source driving these outflows.

The nomenclature for velocity and groupings of data in the present study are the same as in Paper III. Radial velocities are given in \kms\ in the Heliocentric system\footnote{Conversion to LSR is done by subtracting 18.1 \kms }.

\subsection{Outline of this paper}
\label{sec:outline}
The observational material used here is described in Section~\ref{sec:observations}. The importance of the motions in 
the \crossing\ is explained in Section~\ref{sec:CoreMotions}, while the new properties of the HH~269 flow are described in Section~\ref{sec:HH269}. The 3-D positions and motions of the HH~269 Groups are presented in Section~\ref{sec:3DHH269} and their origin in Section~\ref{sec:Origin}. A nearby un-aligned series of shocks is described in Section~\ref{sec:502shocks}. The episodic nature of the HH~269 flow is discussed in Section~\ref{sec:Outflows}. The results of this paper are related to the earlier papers in this series are presented in Section~\ref{sec:discussion}. 

\section{Observations} \label{sec:obs}
\label{sec:observations}
The spectroscopic data we use in this study are the same as in Paper III of this series, and we have made even more extensive use of archival Hubble Space Telescope (HST) images. Again, we draw on the high-spectral-resolution "Spectroscopic Atlas of Orion Spectra" of \citet{gar08}, compiled from a series of north-south slit spectra at intervals of 2\arcsec\ and a velocity resolution of 10 \kms .

\section{The important role of the Orion-S Crossing} 
\label{sec:CoreMotions}

The Orion-S Crossing (henceforth, the \crossing) is a 30\arcsec\ diameter region centered at 5:35:13.95 -5:23:49.2 In Papers II and III we established that this is the region of origin of changes in the foreground Nearer Ionized Layer (NIL) and the ionized surface of the Orion-S embedded molecular cloud.  It is a high (closer to the observer) region on the NE edge of the \cloud .

 \begin{figure*}
\includegraphics
[width=7.5in]
 {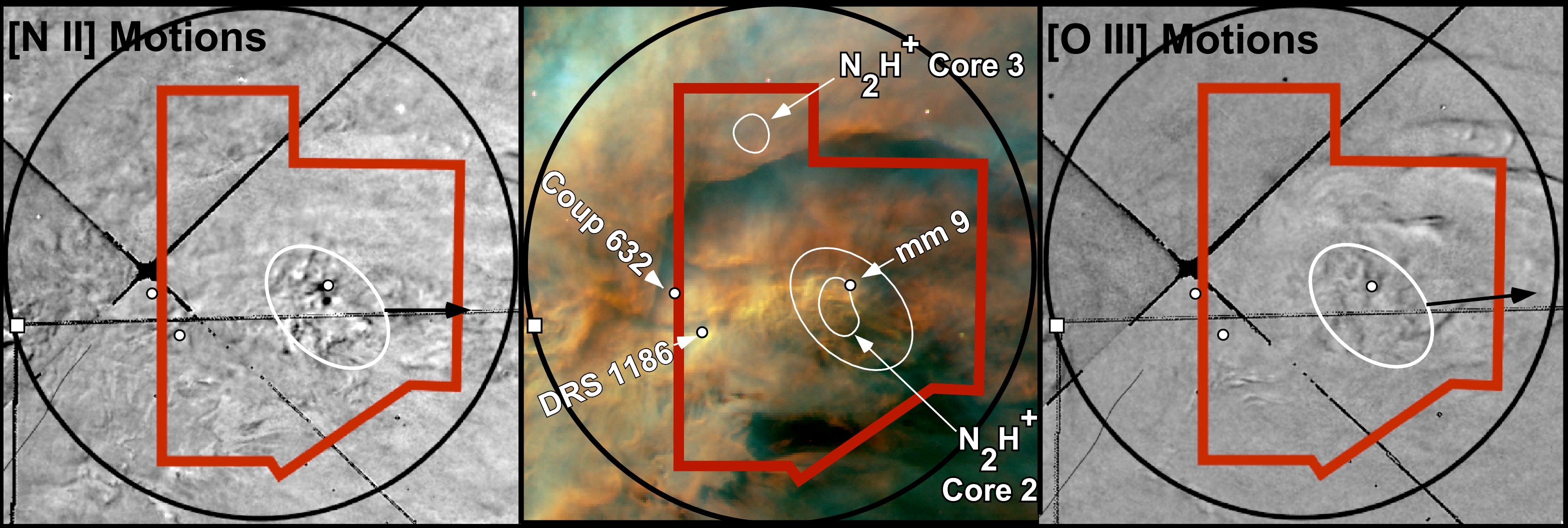}
\caption{These three 30\arcsec $\times$ 30\arcsec\ panels are centered on the \crossing\ (the large black circle). The center panel is made from HST camera WFC3 images with a 
scale of 0\farcs04 pixels (HST program GO 12543, \citep{ode15}), The color coding is Red-\nii, Green-\Ha\ and Blue-\oiii . The other panels depict motions, the left in \nii\ and the right in \oiii. They are the ratio of early over 
late images, so motion is indicated by the dark edge outside of a bright edge as described in \citet{ode15}. The red line gives the boundaries of the \core , where the \Vlowoiii\ and \Vlongoiii\ signals are comparable. The long thin black lines are artifacts at the edges of the earlier images. The three small filled circles are at the positions of the compact sources within the \crossing\ that lie near the central axis of the HH~269 flow. 
The white ellipse encloses the features for which tangential motions could be determined (the \CrGrp ). The black arrows indicate the average tangential velocities for the group (from Table~\ref{tab:Velocities} for both \nii\ (\Vtnii\ = 25$\pm$12 \kms , PA = 270$\pm$10\arcdeg ) and \oiii\ (\Vtoiii\ = 32$\pm$12 \kms , PA = 276$\pm$35\arcdeg ). The irregular white lines designate the central contours of the \mol\ Cores 2 (south) and 3 (north) from \citet{teng20}, Section~\ref{sec:mm9}. The square indicates the position from which the HH~1132 east-jet emerges, as discussed in Section~\ref{sec:PandD}. 
}
\label{fig:Three}
\end{figure*}

Within the \crossing\ we define as the \core\ the region where the three Profiles discussed in PaperIII show that the signals from the \Vlongoiii\ (\oiii\ 500.7 nm emission), and \Vshortoiii\ emission abruptly become comparable. The former isa longer wavelength component of the deconvolved line profile that is usually identified with the main ionization front (MIF) while the latter is a shorter wavelength component  usually identified with foreground lower velocity material. This common region is shown in red outline in Figure~\ref{fig:Three}. 
The \core\ agrees in position with the areas in Paper III where we found that the signal ratios (\Sshortoiii /\Slongoiii) are anomalously strong. usually from the MIF.

Within the \core\ we found blue-shifted components not found in nearby spectra. Their velocities are given in Table~\ref{tab:Velocities} along with velocities from HH~269 features identified in Figure~\ref{fig:HH269}.  Since the spectra were relatively wide, the location of these components are most likely to be in the central region covered by the three profiles. This region was included in the tangential velocity study of \citet{ode15}, which found 
multiple high tangential features. The outer panels of Figure~\ref{fig:Three} show two tangential velocity motion images (first epoch image over second epoch image). The light-line ellipse encloses the region where moving objects were found (seven in \nii\ and two in \oiii). 

Three compact sources in the Crossing and lying near the axis of HH~269 are shown in all panels of Figure~\ref{fig:Three}:  \coup\ \citet{get05} (5:35:14.40 -5:23:50.9),
\drs\  \citet{Dar} (5:35:14.29 -5:23:53.1), and \mm\ \citet{eisner06} (5:35:13.72 -5:23:50.6). 

Using both the average radial and tangential velocities we calculated the spatial motions presented in Table~\ref{tab:3dVelocities}. The high velocity features arising from the Crossing are moving at about
65 \kms\ at an angle $\Theta$ $\simeq$ 60\arcdeg\ from the plane-of-the-sky (toward the observer)  toward Position Angle (PA) $\simeq$ 276\arcdeg  . 

\section{Properties of the HH~269 series of shocks}
\label{sec:HH269}

The series of features associated with HH~269 are among the largest structures within the Huygens Region. Their two brightest components were first studied in detail by \citet{walter} and subsequently designated as HH~269-East and HH~269-West \citep{bom} (henceforth, \east\ and \west\ in this study). The \Htwo\ study of \citet{stanke} 
 covered the full Extended Orion Nebula (EON) and \citet{ode15} assigned Stanke's more distant feature 2-4 as part of HH~269 on the basis of its orientation along the axis of the \east -\west\ features. \citet{ode15} added multiple small moving features to the west of the \west\ component.
 
\subsection{Two new components of HH~269}
\label{sec:TwoNew}
 
\begin{figure*}
\includegraphics
[width=7.0in]
{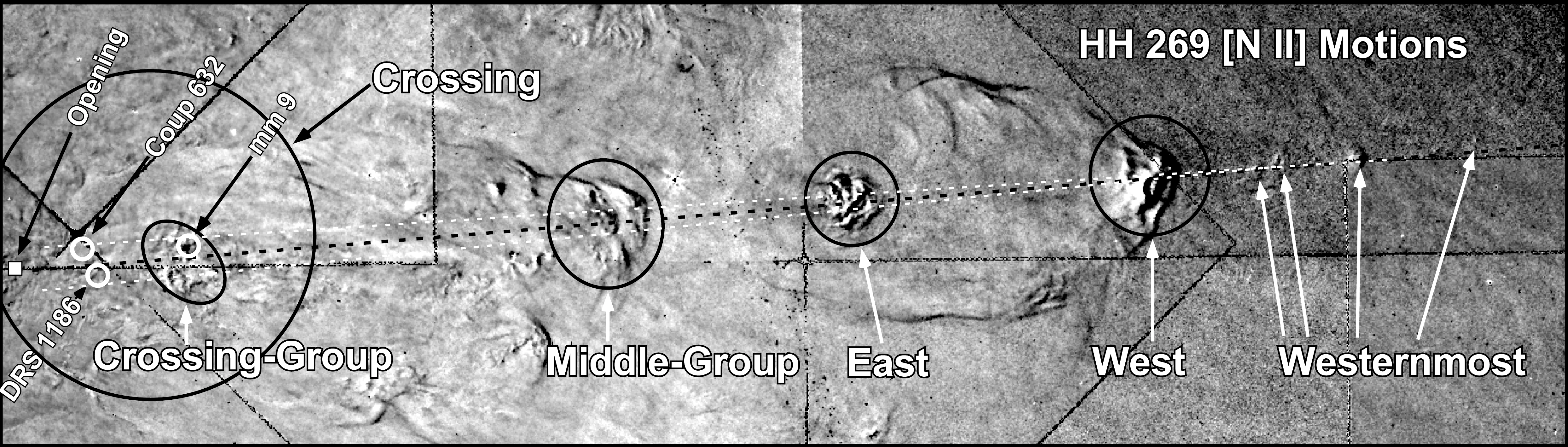}
\caption{This 143.2\arcsec $\times$40.6\arcsec\ F658N \nii\ motions image centered at 5:35:10.0 -5:23:50 shows the \east\ and \west\  components of HH~269 designated in \citet{bom}, the Westernmost components designated in \citet{ode15}, the \midGrp\ components studied in \citet{ode15} that we now designate as part of the HH~269 system, and the \CrGrp\ component highlighted in this study. The dashed black line indicates the common axis of the components at PA = 275\arcdeg\ and the dashed white lines $\pm$1\arcdeg\ uncertainty in that value.  The \Htwo\ knot 2-4 \citep{stanke} lies 91\arcsec\ at PA = 277\arcdeg\ from the most westerly Westernmost marked component. The positions of \coup , \drs , and \mm\ are shown as open circles and the filled white square indicates the position from which the east-moving HH~1132 east-jet emerges, as discussed in Section~\ref{sec:PandD}. The large circle on the east end indicates the \crossing\ from Figure~\ref{fig:Three}. 
}
\label{fig:HH269}
\end{figure*}

We now add two groupings of shocks (the \midGrp\ and the \CrGrp ) to extend the series of components to the east (the latter addition was suggested in \citet{ode15}), as shown in Figure~\ref{fig:HH269}. A major feature within the \crossing\ is the low ionization ''squiggly" feature called the West-Jet in \citet{ode15}. In Paper III we established that  the West-Jet name is not a good description because only the west end of it has a detectable motion, while a \nii\ bright feature 
at 5:35:13.90 -5:23:52 is the highly tilted side of a stationary escarpment, facing north. The latter feature is called the \ledge\ in Paper III and all of the West-Jet is now called the \extledge , which is a mix of moving and stationary features. All of these lie along an axis of PA = 275\arcdeg. The observed and derived characteristics of the components are 
 presented in Tables~\ref{tab:Velocities}, \ref{tab:3dVelocities}, and \ref{tab:positions}. 
 
 \begin{table*}
\caption{Tangential and Radial Velocities of HH 269 Components$^{a}$}
\label{tab:Velocities}
\begin{tabular}{lllllll}
\hline
\hline
\colhead{Group}&{~~\Vtnii}&{~~\PAnii}&{~~\Vnii }&            {~~\Vtoiii  }&           {~~\PAoiii}&             {\Voiii}\\
\hline
\CrGrp\     &   25$\pm$12  &270$\pm$10 & -11$\pm$3    & 32$\pm$12 & 276$\pm$35  & -30$\pm$3\\
\midGrp $^{b}$          &  31$\pm$11  &273$\pm$16 &  -9$\pm$3     & 36$\pm$19 & 262$\pm$14& - 40$\pm$1 \\
\east $^{c}$         &       42$\pm$9  &  277$\pm$13 & -13$\pm$1   &   ---              &    ---              & ---\\
\west\         &       73$\pm$16 & 270$\pm$11 & -23$\pm$2    & 52$\pm$19  &288$\pm$7  &   -17$\pm$3\\
\hline
\end{tabular}\\
~$^{a}$Position Angles (PA) in \arcdeg\ are from \citet{ode15} except where measured in this study and tangential velocities are calculated using an assumed distance of 388 pc \citep{mk17}. The positions of non-Crossing Groups are indicated in Figure~\ref{fig:HH269} and are given in Table~\ref{tab:positions}.

$^{b}$ \Vnii\ is from Large Sample 80,-30 in Paper II.

$^{c}$ \Vnii\ is from \citet{walter}.
\end{table*}

\begin{table}
\caption{HH 269 Components 3-D Velocities and Angles}
\label{tab:3dVelocities}
\begin{tabular}{lccc}
\hline
\hline
\colhead{Designation} &
\colhead{Line} &
\colhead{\Vd $^{a}$} &
\colhead{$\Theta$$^{a}$}\\
\hline
 \CrGrp\ &\nii\ &69$\pm$12 &57$\pm$7\arcdeg \\
  ~~~~"   &  \oiii\ &65$\pm$12 &61$\pm$9\arcdeg\\
\midGrp\  & \nii\ &46$\pm$12   &49$\pm$10\arcdeg\\
 ~~~~" &  \oiii\ &76$\pm$20  &62$\pm$11\arcdeg\\                                 
 \east\       &  \nii\ &58$\pm$10 &44$\pm$6\arcdeg\\                               
\west\     &  \nii\ &88$\pm$15    &34$\pm$6\arcdeg\\
 ~~~"         &  \oiii\ &68$\pm$20  &40$\pm$14\arcdeg\\                                               
Average  &\nii\ $\&$ \oiii &66$\pm$14 &48$\pm$10\arcdeg\\
\hline
\end{tabular}\\
~$^{a}$V$\rm_{3-D}$ is the spatial velocity of the Group in \kms , $\Theta$ is the angle of the velocity vector out of the plane-of-the-sky and toward the observer. The assumed radial velocity of the \cloud\ was 27 \kms.  The average was calculated giving triple weight to the much more numerous \nii\ motions. 
\end{table}

\begin{table}
\caption{Positions$^{a}$ of HH 269 Components}
\label{tab:positions}
\begin{tabular}{lcc} 
\hline
\hline
\colhead{Designation} &
\colhead{RA} &
\colhead{DEC}\\
\hline
\CrGrp\ &5:35:13.71 & -5:23:52\\
\midGrp\ & 5:35:11.08 & -5:23:49\\
\east\  &    5:35:09.65 & -5:23:47\\
\west\ &       5:35:07.77 & -5:23:46\\
Westernmost& 5:35:07.12 & -5:23:44\\
Westernmost& 5:35:06.98 & -5:23:43\\
Westernmost& 5:35:06.46 & -5:23:43\\
Westernmost& 5:35:05.75 & -5:23:41\\
Stanke 2-4 &5:34:59.8 & -5:23:30\\
\hline
\end{tabular}\\
~$^{a}$ {2000.0}
\end{table} 

\subsection{3-D positions of the HH 269 components}
\label{sec:3DHH269}

\Vt\ (motion in the plane-of-the-sky) and \Vr\ (radial velocity)  have been determined for the sequence of HH 269 components \CrGrp ---\midGrp ---\east ---\west\ (Table~\ref{tab:Velocities}).
This allows the calculation of the spatial velocity (\Vd) and the angle ($\Theta$) with respect to the plane-of-the-sky with the results shown in Table~\ref{tab:3dVelocities}. Having the separations in the plane-of-the-sky (Table~\ref{tab:positions})
and the velocity vectors, one can calculate the relative positions along the line-of-sight. Using these results we have calculated the positions of each component with respect to the \CrGrp\ (Table~\ref{tab:3dPositions}), where we see that the HH~269 sequence is highly tilted with respect to the plane-of-the-sky. 

We conclude that the \crossing\ is where the axis of the HH~269 sequence emerges from behind the surface of the \cloud . The source of the components must be at or east of the \CrGrp .

\begin{table}
\caption{3-D Positions of HH~269 Components}
\label{tab:3dPositions}
\begin{tabular}{lccc}
\hline
\hline
\colhead{Component }& \colhead{$\Delta$Tangential$^{a}$ } & \colhead{Adopted $\Theta$} & \colhead{$\Delta$D$^{a}$}\\
\hline
\CrGrp\                      &    0.0   &                                          59\arcdeg                     & 0\\
\midGrp\                        &    0.065&                                          56\arcdeg                     & -0.11\\
\east\                             &   0.105&                                           44\arcdeg                     &-0.17\\
\west\                           &   0.157&                                            ---                                 &-0.22\\
\hline
\end{tabular}\\
~$^{a}$Distances are in pc relative to the \CrGrp\ in the plane-of-the-sky ($\Delta$Tangential) and along the line of sight ($\Delta$D, where negative numbers are toward the observer). \end{table}
 
 \section{Origin of the HH~269 components}
 \label{sec:Origin}
  
 Two basic approaches are used to determine the location of the source of the HH~269 sequence, proximity to the axis of the HH~269 components, and locations suggested by V$\rm_{3-D}$. We will see that these suggest two likely sources.
 
\subsection{Origin from positions and directions}
\label{sec:PandD}
 The dashed black line with PA = 275\arcdeg\ in Figure \ref{fig:HH269} shows that the axis of the HH~269 components
 pass near three compact sources (from the west \mm , \drs , and \coup ) within the Dark Arc (the dark feature along the top of the \crossing ). 
 A plausible uncertainty of the PA of $\pm$1\arcdeg\ would correspond to an uncertainty in Declination of $\pm$1\farcs8 in the \crossing , thus allowing an alignment of the HH~269 axis with any of the three compact sources. At a greater distance the alignment passes close to AC Ori.
 
\subsubsection{\mm\ and an associated \mol\ clump}
\label{sec:mm9}
 \mm\ is located in the NW portion of the elliptical region defining the shocks associated with the \CrGrp .
 It is located slightly offset within the \CrGrp\ but along the axis of the HH~269 flow if that axis has the allowable error of 1\arcdeg. It was measured at 3 mm as a $>$5 $\sigma$ detection by \citet{eisner06}, who note that it has no near infrared counterpart. 

 In addition, \mm\ lies within Core 2 found by \citet{teng20} in their 5\arcsec\ resolution map in \mol\ emission in the J = 3--2 line along a north-south feature within the background Orion Molecular Cloud. Two of the \citet{teng20} cores are shown in Figure~\ref{fig:Three}. Core 2's heliocentric velocity is about 24.6 \kms. 
 More recently \citet{hacar} mapped the same region with 10\arcsec\ resolution in the \mol\ J = 7--6 transition, an emission-line selectively coming from very high density gas n(\Htwo ) $>$ 10$^{7}$ \cmq. In that study they established a strong correlation of high density \mol\ knots and very young stars, which argues that this region contains a potential source of the material causing the HH~269 shocks.
 Although \citet{teng20} argued that the Core 2 lies in the background Orion Molecular Cloud, \citet{hacar} place it within the \cloud .

 \subsubsection{\drs}
 \label{sec:drs1186} 

 \drs\ lies closest to the HH~269 axis of 275\arcdeg . Little is known about it, although the discovery paper \citep{zap04} shows the source to have a micro-jet of about 0\farcs5 extending to the NW, not at all close to the axis of the HH~269 features. This makes it an unlikely source. Nearby there is a west oriented shock at 5:35:14.3 -5:23:51 with \Vtnii\ = 8 \kms\ but is at the limit of detectability. The shock's position gives no support for association of \drs\ with HH~269. 

\subsubsection{\coup}
\label{sec:coup}

\coup\ lies on the northern boundary of the axis of HH~269. Appendix A9 of \citet{riv13} summarizes well the characteristics of \coup , although they incorrectly assign it as the source of HH~529 \citet{ode15}. It is seen in X-rays, through the infrared, and in short wavelength radio radiation. 

\coup\ was identified in \citet{ode15} as the source of the HH~1132 east-jet that emerges with PA = 107\arcdeg\ from within the Orion-S Cloud at the position shown with a square in Figures~\ref{fig:Three} and \ref{fig:HH269}. The emergence position is quite obvious in time sequence F656N \Ha\ images \citep{ode15} rendered as a movie.
The axis of this east-jet points exactly at \coup\ with an opening-\coup\ 
separation of 6\farcs5.

This means that \coup\ is the likely source of one collimated outflow (HH~1132 east-jet), but linking it to HH~269 requires that another jet be pointed toward PA = 275\arcdeg. A bipolar system would require the west-jet to be pointed toward PA = 287\arcdeg. A compounding problem to association of the HH~1132 east-jet and the HH~269 shocks is that both are blue-shifted. The east-moving HH~1132 moves at a spatial velocity of 116 \kms\ with $\Theta$ =  32\arcdeg \citep{ode15} while the \CrGrp\ shocks have 
PA = 275\arcdeg, $\Theta$=59$\pm$10\arcdeg , and space velocity 67$\pm$12 \kms . 

A strong selection effect exists for finding blue-shifted components in the Huygens Region because the red-shifted component would be headed toward the MIF and disappearance behind the PDR. Likewise, a red-shifted jet originating behind the PDR will not emerge into the low extinction ionized gas. A summary of outflows \citet{ode15} finds 27 mono-polar flows (all blue-shifted), eight bipolar flows (none with a well established red-shift), and seven multi-polar flows (all blue-shifted). Therefore, there is plenty of evidence for non-bipolar flows in the Huygens Region and \coup\ may be such a source. 

If \coup\ is the source, then it lies within the \cloud .
The separation of the \CrGrp\ of moving features from \coup\ is 10\farcs1 (0.0188 pc). 
Adopting that object as the source and using $\Theta$ = 59\arcdeg, puts \coup\ (0.031 pc) beyond the plane containing the \CrGrp\ shocks. 
The separation in the plane-of-the-sky of the \CrGrp\ shocks and the SE-NW Transition established in Paper II is 23\arcsec\ (0.0427 pc). 
This means that \coup\ is still within the Cloud. 
If it is also the source of the HH~1132 east-jet, then this interpretation is fully consistent with the break-out of that feature.

\section{Two series of shocks within the Crossing that are not related to HH~269}
\label{sec:502shocks} 

\begin{figure}
\includegraphics
[width=3.0in]
{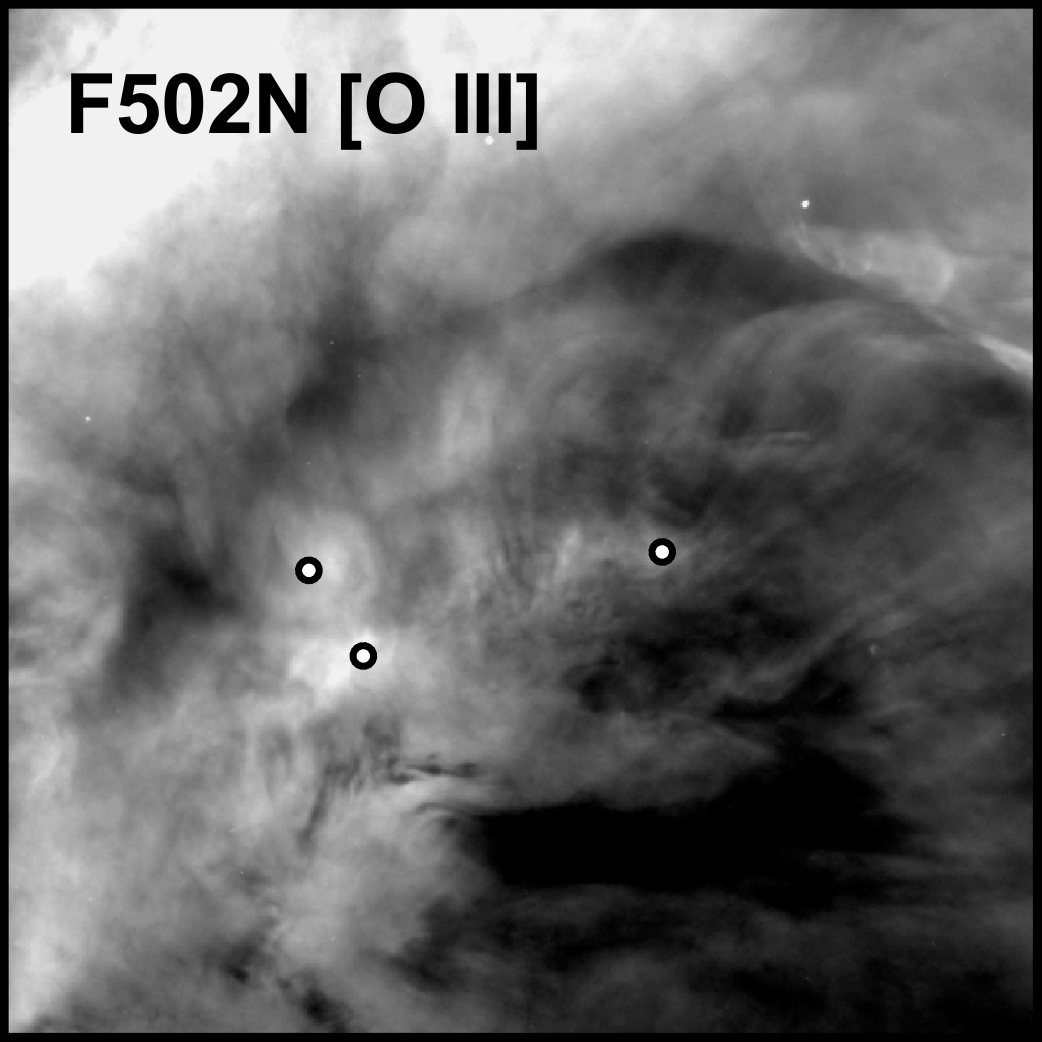}
\caption{The same Field-of-View as Figure~\ref{fig:Three} except now showing the filter WFC3 F502N \oiii\ image. The display is adjusted to best show the faint 
arcs north of source mm 9 but inside the Dark Arc. Their motions were reported in \citet{ode15}. 
}
\label{fig:Crossing502}
\end{figure}

There are large-scale shocks of note falling within the \crossing\ that have no relation to HH~269.
In the tangential velocities study of \citet{ode15} using HST WFC3 and WFPC2 images, a series of concentric strongly curved arcs was found in \oiii\ motions images in the \crossing , as shown in Figure~11 of that paper. These lie between \mm\ and the Dark Arc and were attributed to outflow from an undetected source a few arcseconds north of \mm . We show in Figure~\ref{fig:Crossing502} that these features are also seen in the superior WFC3 \oiii\ images adjusted in brightness and contrast to best display this area. We now see that they are not a circular motion away from 
an empty position designated as the Blank-West in \citet{ode15}. Instead, they are the lead features in a series of bow-shock shapes driven by a more distant source to the SSW.  At 
7\farcs5 west of their crests a series of three partial shocks is also seen. Their more northerly orientation indicates a different source than that producing the former group of bow-shocks. If they do have the same source, their axes intersect well south of the \crossing .

\section{Most of the HH~269 components are formed from episodic outflows}
\label{sec:Outflows}

The series of well-defined bow-shocks that we see in the \midGrp , \east , and \west\ components argue for the driving jet being episodic because each period of flow would produce a single shock or a set of tightly grouped shocks.  The \Vt\ and separations of the four main components indicate the time intervals
between outflows:
 \CrGrp\ and \midGrp , 1900 yrs; \midGrp\ and \east , 2500 yrs; and \east\ and \west , 2600 yrs. These time intervals are similar to those found in other outflows in the Huygens Region \citep{ode08a}. 


The HH~269 \CrGrp\ component lacks the well-defined bow-shocks seen in the \midGrp , \east , and \west\ 
components. This is consistent with this component not representing a single outflow, rather, that it is 
where a submerged outflow passes through the near side of the Orion-S Cloud.  

\section{Discussion and Conclusions}
\label{sec:discussion}

In Paper III we established that the \crossing\ is centered on a local rise in the nearer side of the \cloud , which then flattens to the SW. Now we see that the \CrGrp , the first visual components of HH~269, appears there. This suggests that a change in the local topography has determined where the flow producing the \CrGrp\ breaks out. 

This conclusion is reinforced by the fact that the E-W line along which the HH~269 features appear is also where the \oiii\ emission changes from domination 
by a component clearly near the MIF to lower velocity components associated with the NIL, a behavior that extends to the SW. 

$\bullet$ Taken together, these points mean that the structure of the nearer ionized layer of the \cloud\ influences where one can see the results of the HH~269 flow, and this flow does not influence the structure.

$\bullet$ Two young stars are candidate sources of the collimated outflow that drives HH~269-the \mm\ source associated with \mol\ Clump 2 and the nearby \coup .

$\bullet$ At 10\farcs1 the obscured star \coup\ has all the properties of a young star that produces jets. It has already been 
identified as the source of HH~1132, which is moving toward 108\arcdeg\ or slightly less \citep{ode15}. The reciprocal of its motion is 288\arcdeg ,
producing a poor match to HH~269's PA = 275$\pm$1\arcdeg. More troubling is that both HH~1132 and HH~269 are clearly blue-shifted (\Vrn =-40 \kms \citep{ode15}  and -14$\pm$5 \kms , respectively). However, multi-polar outflows in the Huygens have been found \citep{ode15}.  If this is the source, it lies in the western portion of the \cloud .

$\bullet$ \mm\ is located within the high-density \mol\ Clump 2, which is a likely source of new star formation, and must lie close to the surface of the ionized layer of the \cloud\ that faces the observer. Given its high extinction, it is likely to be within or beyond the underlying PDR.

$\bullet$ The spatial separations and velocities of the components of  HH~269 argue that these are the result of intermittent jet activity at intervals of about 
1900 to 2600 years.

 \section*{acknowledgements}
 
The observational data were obtained from observations with the NASA/ESA Hubble Space Telescope,
obtained at the Space Telescope Science Institute (GO 12543), which is operated by
the Association of Universities for Research in Astronomy, Inc., under
NASA Contract No. NAS 5-26555; the Kitt Peak National Observatory and the Cerro Tololo Interamerican Observatory operated by the Association of Universities for Research in Astronomy, Inc., under cooperative agreement with the National Science Foundation; and the San Pedro M\'artir Observatory operated by the Universidad Nacional Aut\'onoma de M\'exico. 
We have made extensive use of the SIMBAD data base, operated at CDS, Strasbourg, France and its mirror site at Harvard University, and to NASA's Astrophysics Data System Bibliographic Services. 
GJF acknowledges support by NSF (1816537, 1910687), NASA (ATP 17-ATP17-0141), and STScI (HST-AR- 15018).

\clearpage

\end{document}